\documentclass{aa} 
\usepackage{graphicx}
\usepackage{txfonts} 
\usepackage{subfigure} 
\usepackage{soul} 
\usepackage{color} 
\usepackage{here} 
\usepackage{natbib}
\bibpunct{(}{)}{;}{a}{}{,}   

\newcommand{\scsc}[1]{{\scriptscriptstyle{#1}}} 
\newcommand{\dgr}{\hbox{$^\circ$}}   

\newcommand{\GHz}{\hbox{$\,$GHz}}
\newcommand{\nuIF}{\hbox{\ensuremath{\nu_{\mathrm{\scsc{IF}}}}}}
\newcommand{\nuRF}{\hbox{\ensuremath{\nu_{\mathrm{\scsc{RF}}}}}}
\newcommand{\nuLO}{\hbox{\ensuremath{\nu_{\mathrm{\scsc{LO}}}}}}
\newcommand{\phiLO}{\hbox{\ensuremath{\phi_{\mathrm{\scsc{LO}}}}}}
\newcommand{\Dnu}{\hbox{\ensuremath{\Delta \nu}}}

\newcommand{\taui}{\hbox{\ensuremath{\tau_\mathrm{i}}}}
\newcommand{\taug}{\hbox{\ensuremath{\tau_\mathrm{g}}}}
\newcommand{\taupc}{\hbox{\ensuremath{\tau_\mathrm{pc}}}}
\newcommand{\taulg}{\hbox{\ensuremath{\tau_\mathrm{lg}}}}
\newcommand{\Dtau}{\hbox{\ensuremath{\Delta\tau}}}

\newcommand{\Rx}{\hbox{\ensuremath{R_{\mathrm{x}}}}}
\newcommand{\Ffull}{\hbox{\ensuremath{f_{\mathrm{fr}}}}}
\newcommand{\Ffr}{\hbox{\ensuremath{f_{\mathrm{mfr}}}}}

\newcommand{\imagj}{\ensuremath{{\rm j}}}
\newcommand{\paperone}{H07}

\newcommand{\OmegaE}{\ensuremath{\Omega_{\scsc{\rm E}}}}
\newcommand{\Tpcb}{\ensuremath{T_{\scsc{\rm pc}}}}
\newcommand{\nuf}{\ensuremath{\nu_{\rm f}}}
\newcommand{\revd}[1]{#1} 
\begin{document}
   \title{A spectral synthesis method to suppress aliasing\\
          and calibrate for delay errors in Fourier transform correlators}
          \titlerunning{A method to suppress aliasing and calibrate
          for dealay errors} 

   \subtitle{}

   \author{Tak Kaneko\inst{}
          \and
	  Keith Grainge\inst{}
	}

   \offprints{T.~Kaneko}

   \institute{Cavendish Laboratory, Cambridge University, 
	      Cambridge CB3 0HE, United Kingdom.\\
	      \email{tk229@mrao.cam.ac.uk}
	     }

   \date{Received MMMMMMMM DD, YYYY; accepted June 30, 2008}

  \abstract
  {Fourier transform (or lag)
  correlators in radio interferometers can serve as an efficient means
  of synthesising spectral channels. However aliasing corrupts the
  edge channels so they usually have to be excluded from the data
  set. In systems with around 10 channels, the loss in sensitivity can
  be significant. In addition, the low level of residual aliasing in
  the remaining channels may cause systematic errors. Moreover, delay 
  errors have been widely reported in
  implementations of broadband analogue correlators and simulations
  have shown that delay errors exasperate the effects of aliasing.}
   {We describe a software-based approach that suppresses aliasing by
  oversampling the cross-correlation function. This method can be
  applied to interferometers with individually-tracking antennas
  equipped with a discrete path compensator system. It is based on the
  well-known property of interferometers where the drift scan response
  is the Fourier transform of the source's band-limited spectrum.}
   {In this paper, we simulate a single baseline interferometer, both
  for a real and a complex correlator. \revd{Fringe-rotation usually 
  compensates for the phase of the fringes to bring the phase centre 
  in line with the tracking centre. Instead, a modified fringe-rotation
  is applied. This enables an oversampled cross-correlation function to be  
  reconstructed by gathering successive time samples.}}
   {Simulations show that the oversampling method can synthesise the 
  cross-power spectrum while avoiding aliasing and works robustly 
  in the presence
  of noise. An important side benefit is that it naturally accounts for 
  delay errors in the correlator and the resulting spectral channels
  are regularly gridded}
   {}


   \maketitle
%


\section{Introduction}\label{sec:intro}

\begin{figure}[hbtp]
\centering
  \mbox{\subfigure[Real correlator]{
  \includegraphics[width=3.5in,trim=0 0 0 65,clip=true]{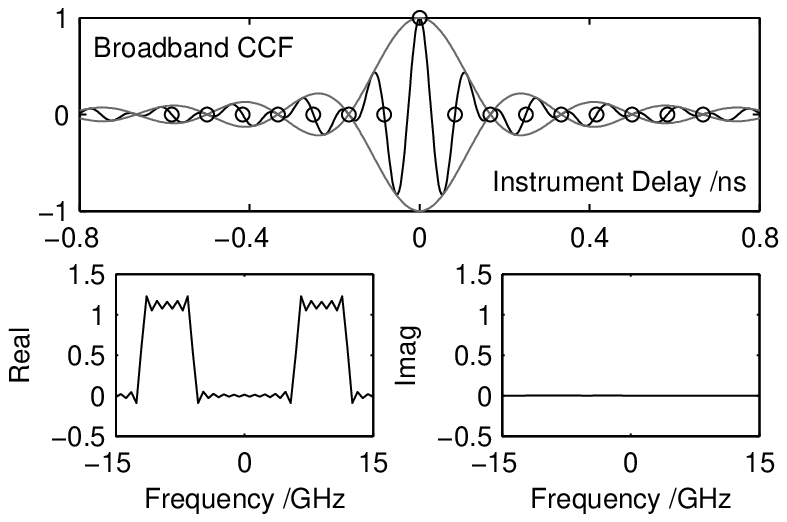} }} 
  \mbox{\subfigure[Complex correlator]{
  \includegraphics[width=3.5in]{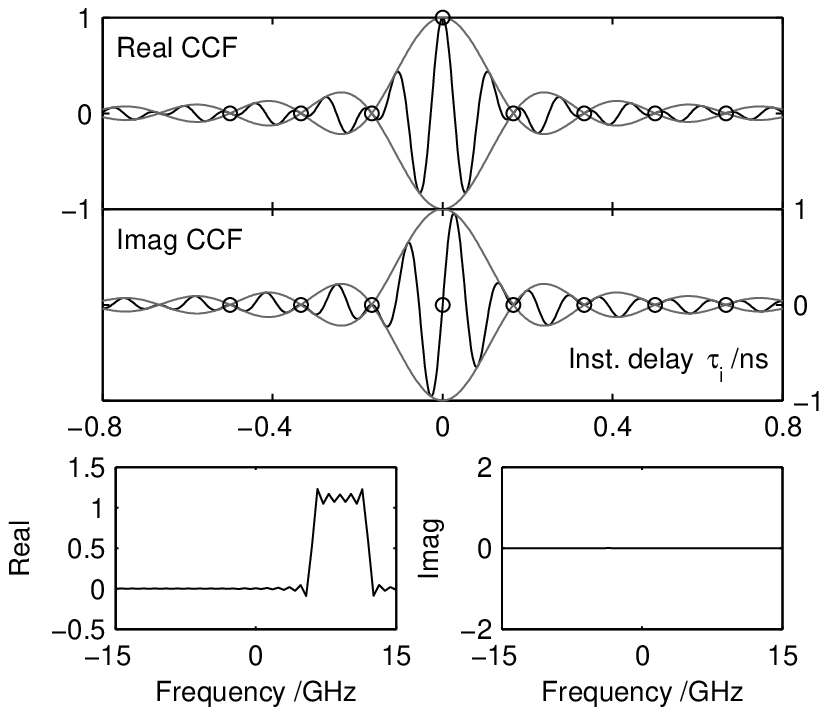} }}
\caption[The real and complex CCF]{{\bf a)} {\it Top plot:} The
  cross-correlation function measured by a real correlator (black) and
  the envelope (grey). This is a graphical representation of the
  cross-correlation function in Eq.~(\ref{eqn:corr-response}) with the
  geometrical delay $\taug = 0$. The open circles indicate the
  position of the detectors in instrument delay $\taui$. {\it Bottom
    plots:} The cross-power spectrum recovered by applying the Fourier
  transform on the cross-correlation function. We have assumed the
  telescope parameters in Table~\ref{tab:tel-param}. {\bf b)} The
  in-phase (real) and quadrature (imaginary) cross-correlation
  functions sampled by a complex correlator. The cross-power spectrum
  recovered in this case is single-sided.}\label{fig:ccf-spec}
\end{figure}

The observing band of radio interferometers frequently needs to be
split into sub-bands, either for spectral line observations or to
reduce the effects of chromatic aberration. Fourier transform
correlators offer an efficient method for dividing the observation
band. In this scheme, the signal from the two arms of the
interferometer are correlated at discrete delay steps, making direct
measurements of the cross-correlation function (see
Fig.~\ref{fig:ccf-spec}). Taking the Fourier transform of the
cross-correlation function gives the complex cross-power spectrum. For
a signal of bandwidth $\Delta\nu$, Nyquist sampling theorem requires
the signal to be sampled at \revd{time intervals of $1/(2\Delta\nu)$}
to avoid aliasing. But the cross-correlation function of a
band-limited signal extends over an infinite delay range. So the
Nyquist sampling theorem holds true only if we sample the
cross-correlation function over an infinite delay range. Clearly this
is not practical so the cross-correlation function is sampled over a
finite range. This results in a recovered signal spectrum with tapered
band edges which will overlap with its images in the spectral
domain. This overlap causes aliasing and will corrupt the recovered
cross-power spectrum.

\begin{figure*}[hbtp]
\centering
  \mbox{\subfigure[The \revd{baseband signal ({0--12\GHz}) under and critically-sampled. The spectra show signs of aliasing. The cross-correlation function was sampled at 16 and 32 delays.}]{
       \includegraphics[width=17cm]{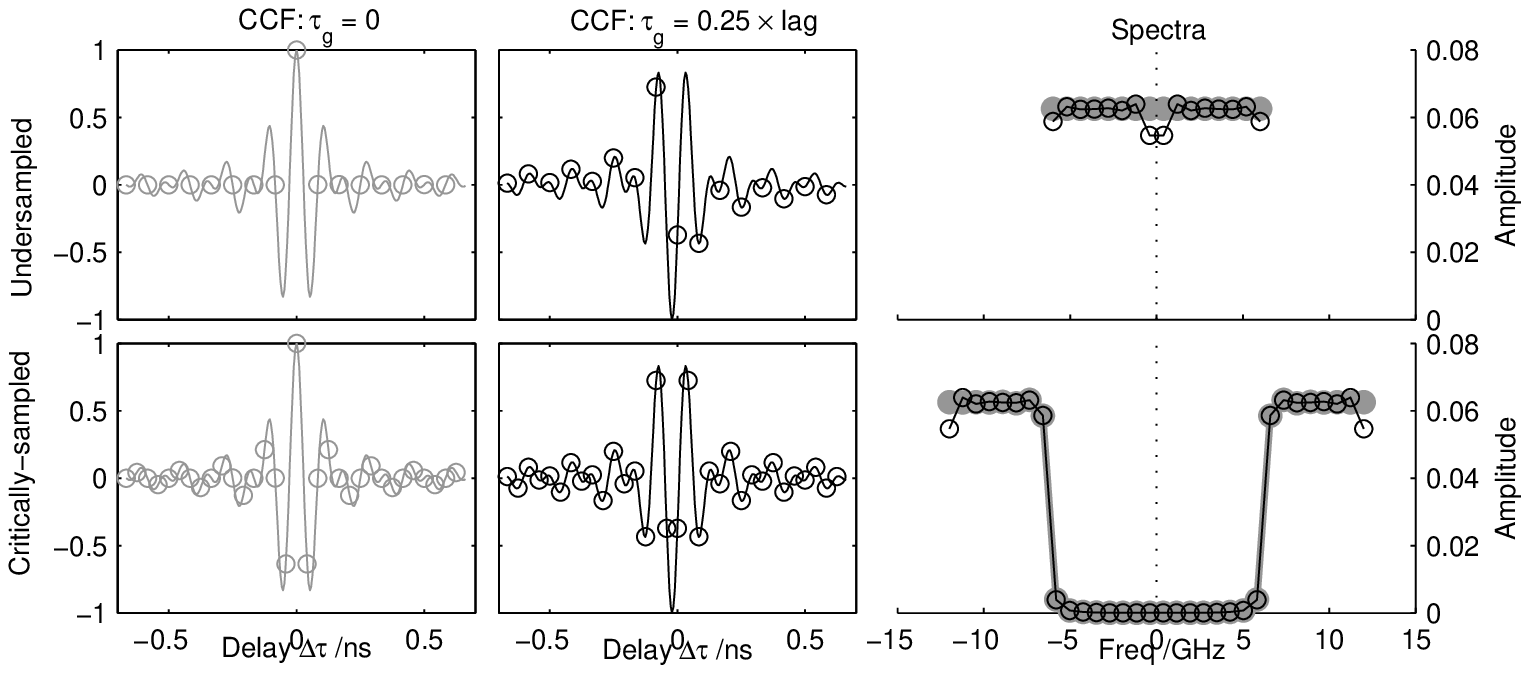}}}
  \mbox{\subfigure[\revd{Oversampling the baseband signal by $2\times$ and $4\times$ critical-sampling reduces aliasing.}]{
       \includegraphics[width=17cm]{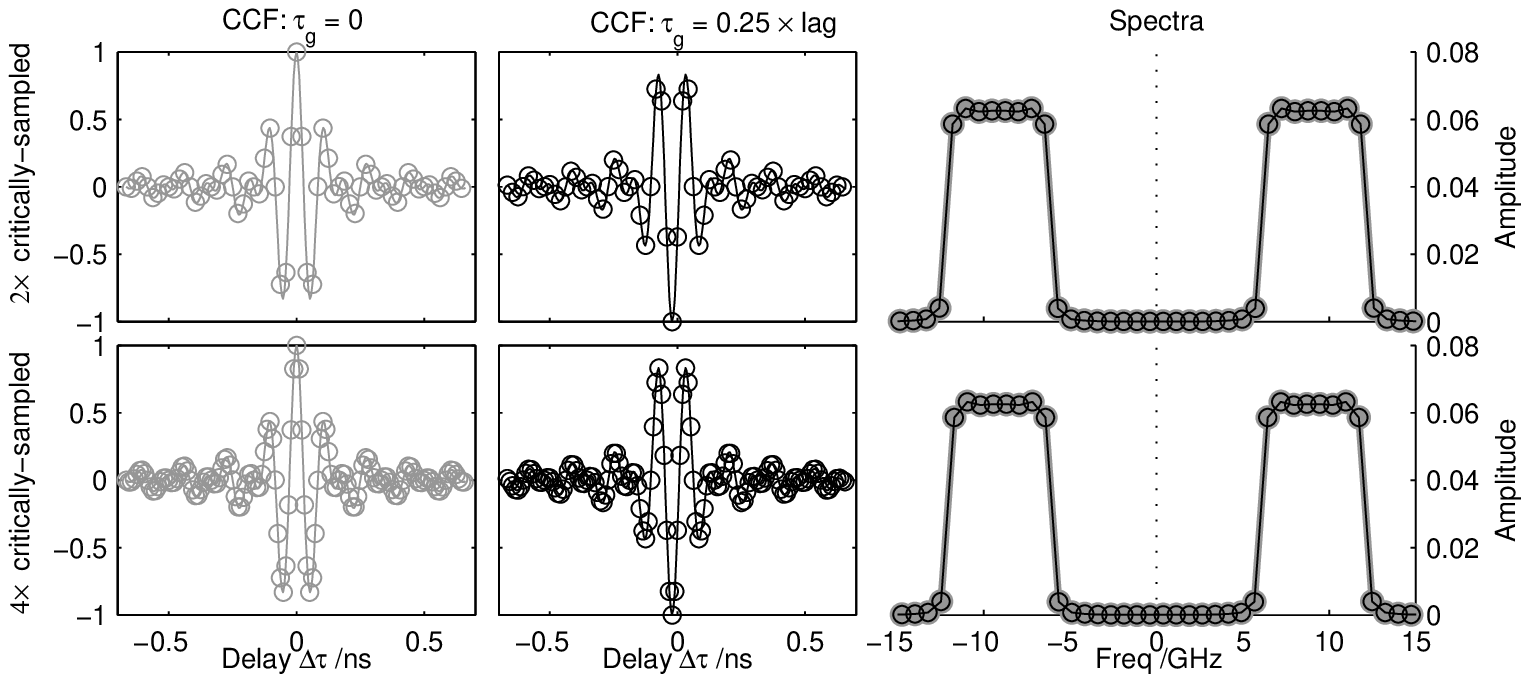}}}
\caption[Critical and oversampling]{The effects of different
  oversampling rates on the recovered spectra. The left-most plots
  show the cross-correlation function \revd{from
    Eq.~(\ref{eqn:corr-response4}) as a function of the residual
    delay {\Dtau}. This is a snapshot when the source transits an
    east-west baseline so the geometrical delay $\taug = 0$ (case~1).}
  The next \revd{set of} plots to the right \revd{(case~2) show the
    cross-correlation function a little later when the centre of the
    envelope has drifted by a quarter of a delay spacing}. The
  measurements at transit \revd{(case~1)} are indicated by the grey
  circles and the measurements later on \revd{(case~2)} are indicated
  by the black circles. The horizontal axis is the delay in {\Dtau},
  which will be defined in Sect.~\ref{sec:real-comp-corr}. The
  right-most plots are the amplitudes calculated from the delay
  measurements. The filled grey marks are at transit \revd{(case~1)}
  and the dark circles are a while later \revd{(case~2)}. The two sets
  of marks should match but it can be seen in {\bf a)} that they do
  not. This mismatch shows up as the temporal modulation seen in
  Fig.~\ref{fig:alias-cycles}. In the upper row of {\bf a)},
  \revd{although the {6\GHz} signal bandwidth} is critically-sampled
  at 16 delays, \revd{the} {0--12\GHz} \revd{baseband signal is
    undersampled.} The two halves of the spectrum are side-by-side.
  The spectrum is aliased at both the high and low-frequency ends.  In
  the lower row of {\bf a)}, the highest frequency component
  ({12\GHz}) is critically-sampled with 32 delays. Now only the upper
  frequency range suffers from aliasing. In {\bf b)}, \revd{the
    baseband signal is oversampled by a factor of 2 and 4 (sampled at
    64 and 128 delays respectively)}. Now the two sets of spectral
  amplitudes match. Oversampling ensures a buffer between the two
  halves of the signal spectrum as well as with the image
  spectra. This reduces the spectral overlap and suppresses
  aliasing. In this case, oversampling the \revd{baseband frequency}
    ({0--12\GHz}) by a factor of 2 is sufficient (\revd{an
    equivalent} sampling frequency of {48\GHz} at 64
  delays). Oversampling above this does not reduce aliasing any
  further. As a minor technical detail, the forms of the spectra in
  {\bf b)} are slightly different from the one in
  Fig.~\ref{fig:ccf-spec}. Here (and also in
  Fig.~\ref{fig:alias-cycles}), we shifted the spectra by half a
  sub-band using the shift theorem of Fourier transforms. This ensures
  that the sub-bands span the desired frequency range and is discussed
  further in \cite{holler2007}. Subsequent simulations will assume
  channel-shifting.}\label{fig:under-vs-oversamp}
\end{figure*}

\revd{As an illustration, the plots in
  Fig.~\ref{fig:under-vs-oversamp} show snapshots of the
  cross-correlation functions and the recovered spectra.} The
left-most plots \revd{(case~1)} show the cross-correlation function
when the source transits an east-west baseline (that is when the path
difference between the two arms of the interferometer is zero). The
circles represent the measurements at discrete delay steps of the
correlator. The next plots to the right \revd{(case~2)} show the
cross-correlation function a short while later. The right-most plots
show the spectrum calculated by taking the Fourier transform of the
discrete measurements. The filled grey circles indicate the spectrum
at transit \revd{(case~1)} and the dark circles are for
\revd{case~2}. In the top row of Fig.~\ref{fig:under-vs-oversamp}a,
the cross-correlation function was \revd{sampled at 16 delays. This
  critically samples the {$\Dnu=6$\GHz} signal bandwidth but
  undersamples the {0--12\GHz} basebandsignal,} so the positive and
negative halves of the {6--12\GHz} bands lie side-by-side. In
\revd{case~1}, the cross-correlation measurements trace out a delta
function and give the characteristic flat spectrum. But in
\revd{case~2}, the amplitudes are perturbed. Ideally, the amplitudes
should \revd{not change} so this points to a fundamental problem. This could
be worked round by oversampling the \revd{baseband signal} at 64 or
more delays as illustrated by Fig.~\ref{fig:under-vs-oversamp}b. Now
the two sets of amplitudes at different times match
perfectly. Oversampling introduces a buffer between the two halves of
the passband and also between their spectral images. This suppresses
aliasing by reducing the overlaps between the signal and image
bands. But clearly, sampling at 64 delays in hardware is not
practical.

\begin{figure}[hbtp]
\centering
  \includegraphics[width=3.5in]{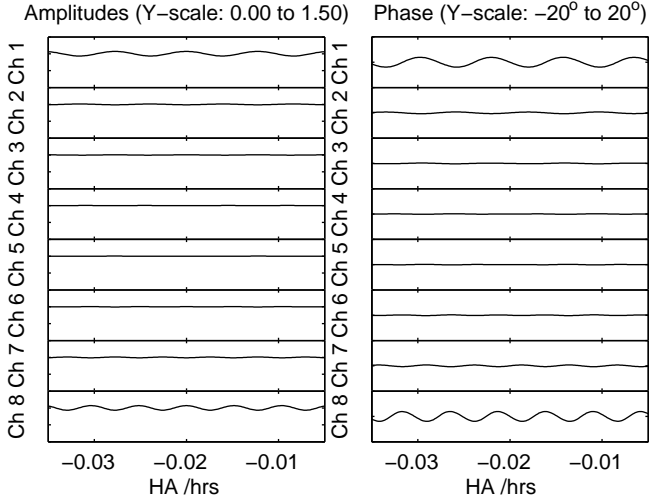}
\caption[Alias cycles]{Simulated time-stream data of the amplitude and 
phase for each spectral channel. The cycles are caused by aliasing and 
are worst in the edge frequency channels. This simulation is for a 
{\it real} correlator with parameters outlined in
Table~\ref{tab:tel-param}. Similar effects are also seen in the 
complex correlator.}\label{fig:alias-cycles}
\end{figure}

\revd{This is manifested in the Fourier Transformed data (the
  spectrum) as temporal modulations in both the amplitude and phase
  (see Fig.~\ref{fig:alias-cycles}).} In aliased signals, the noise
components between the channels will also be correlated so the
individual channels cannot be strictly treated as independent
measurements. These effects are particularly pronounced in the edge
frequency channels as seen in Fig.~\ref{fig:alias-cycles} so these
channels are usually rejected. This may be acceptable in systems with
tens to hundreds of channels. But some correlators, particularly
broadband analogue Fourier transform correlators with channels of
order 10 (for example, \citealt[][ hereafter
  {\paperone}]{li2004,roberts2007,holler2007}), the loss constitutes a
significant portion of the total bandwidth.


In principle, there are three ways to suppress aliasing: 
\begin{enumerate}
\item Increasing the delay range over which the cross-correlation function
is sampled. The spectral channel width will be narrower and the edges
of the passband will be sharper. So the overlap between the signal
spectrum and its images will be narrower. The edge channels will still
have to be rejected but it will be a smaller portion of the whole data.
\item Oversampling the cross-correlation function at finer delay steps
as already illustrated with Fig.~\ref{fig:under-vs-oversamp}.
\item Reducing the bandwidth so that it is not critically-sampled.
\end{enumerate}
Implementing either of the first two modifications in hardware would
be costly and the third approach would lose sensitivity. In
Sect.~\ref{sec:os-method}, we propose a software-based approach to
oversample the cross-correlation function and avoid aliasing. This is
based on the well-known notion that the spectrum of a source can be
obtained from the cross-correlation function measured by a drift
scan. We can reconstruct the cross-correlation function using a
combination of source tracking, path compensation and
fringe-rotation. In Sect.~\ref{sec:simulations}, we will illustrate an
application of this technique with simulations. The system
requirements are: (1) Individually-tracking antennas and (2)
discrete-delay path compensation. We have modelled an analogue
correlator here but the principles could equally apply to digital
correlators. 

\revd{A} number of groups recently reported broadband analogue Fourier
transform correlators suffering from delay errors (for example,
\citealt{harris2001}; {\paperone}; \citealt{roberts2007}). Simulations
by {\paperone} showed that delay errors make the effects of aliasing
worse. Although the delay errors can be calibrated a sample at a time
(recently by \citealt{harris2001}; {\paperone}), the method described
here \revd{will be} a natural way of accounting for delay
errors. This \revd{side benefit} is perhaps as significant as the
alias suppression aspect of the method. The resulting spectral
channels are also regularly gridded at the desired
frequencies. 

However there are a number of issues that must be
considered in a practical system and these are discussed in
Sect.~\ref{sec:practical-issues}. This paper is a follow up to
{\paperone} where we described the development of a broadband
6--12{\GHz} Fourier transform correlator for the Arcminute Microkelvin
Imager \citep[AMI;][]{kaneko2006,zwart2008}. AMI is a new
interferometer designed to survey for clusters of galaxies by
exploiting the Sunyaev-Zel'dovich effect \citep{sunyaev1972}. The
effects of aliasing in the correlator used in this instrument are
discussed further in {\paperone}. Before going into a detailed
discussion of the oversampling method, we will first give a brief
overview of how the data is processed in a conventional
interferometric system.

%

\section{Conventional interferometry}\label{sec:conventional}

\begin{figure}[hbtp]
\centering
  \includegraphics[width=2.5in]{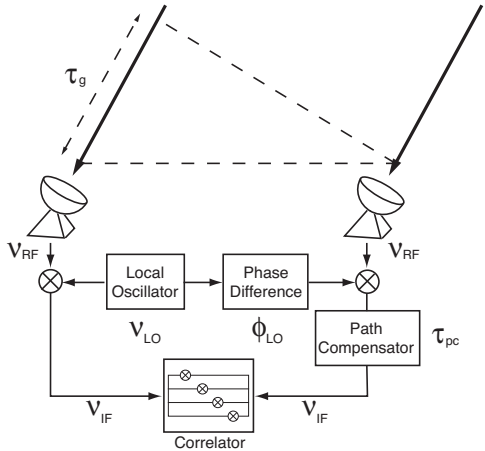}
\caption[Interferometer System]{A schematic of an interferometer
system based on \citet{thompson2001}. The geometric delay ($\taug$)
between a pair of antennas is corrected by the path compensator delay
($\taupc$) in the IF. The correlator samples the cross-correlation
function at a number of delay steps ($\taulg$).}\label{fig:interferom}
\end{figure}

As the antennas of an interferometer track a source, the geometric
path difference ($\taug$) between the two arms needs to be corrected
(Fig.~\ref{fig:interferom}). One approach called path compensation is
to insert discrete lengths of \revd{delay lines ($\taupc$).} Usually, the
signal at the observation frequency {\nuRF} is downconverted by mixing
it with the local osciallator (LO) at {\nuLO}. This converts the
signal to a lower intermediate frequency (IF) at {\nuIF}. The path
compensation is usually inserted in the IF and we will assume that
this is the case. It is possible to eliminate the path compensation
systems all together by co-mounting the antennas. But our spectral
synthesis method cannot be applied to such a system and must have
individually-tracking antennas. Individually-tracking atennas also
give better control on systematic errors because non-astronomical
noise and other contaminating signals can be removed by
fringe-filtering.

An interferometer is a device that measures the cross-correlation
function of the signals received by a pair of antennas. For a
source at the tracking centre, this is given by
\begin{equation}
R_1 = R_0 \frac{\sin \pi \Dnu (\taug - \taui)}{\pi \Dnu (\taug - \taui)}
\cos \bigl[ 2\pi \bigl(\nuLO \taug \pm \nuIF (\taug - \taui)  +
\phiLO \bigr) \bigr]. \label{eqn:corr-response}\\ 
\end{equation}
\revd{The coefficient $R_0$ is proportional to the flux density of the
  point source and also includes factors like the gain of the detector
  circuit.} The plus sign in the $\pm$ term applies to upper side-band
systems and the minus sign is for lower side-band systems. $\phiLO$ is
the phase of the local oscillator and it can be removed by phase
calibration, at a later stage. {\taui} is the total instrument delay
difference in the IF inserted by the path compensator ({\taupc}) and
the differential delays inserted by the correlator lag ({\taulg});
\begin{equation}
\taui = \taupc + \taulg. 
\end{equation}
The cross-correlation function (Eq.~(\ref{eqn:corr-response})) plotted
against the correlator delays {\taulg} (at $\nuIF$) consists of a
cosine term bound by a sinc envelope (see
Fig.~\ref{fig:ccf-spec}a). The Fourier transform of the
cross-correlation function gives the spectrum. So if we can sample the
cross-correlation function sufficiently at regular correlator delays
{\taulg}, \revd{we can measure the spectrum of the source.  This is
the basis of Fourier transform correlators that we will be discussing
in this paper.}

\revd{For brevity, we will define the residual delay at
  each correlator delay as}
\begin{equation}
\Dtau = \taug - \taui = \taug - (\taupc + \taulg). 
\label{eqn:Dtau}
\end{equation}
\revd{We can then re-express Eq.~(\ref{eqn:corr-response}) using {\Dtau}
  and this will be the standard form we will use in the rest of the
  paper;}
\begin{equation}
R_1 = R_0 \frac{\sin \pi \Dnu \Dtau}{\pi \Dnu \Dtau}
\cos \bigl[ 2\pi \bigl(\nuLO \taug \pm \nuIF \Dtau 
+ \phiLO \bigr) \bigr]. \label{eqn:corr-response1}
\end{equation}

\subsection{Real and complex correlators}\label{sec:real-comp-corr}

Broadly speaking, there are two architectures for Fourier transform
correlators. We will call these the real and complex correlators. The
spectral information they give are the same but the way they sample
the cross-correlation function is different. A more detailed
discussion on Fourier transform correlator designs can be found, for
example, in \paperone. The real correlator samples at $2N$ delay steps
and will produce $N$ sub-bands. The complex correlator makes two
measurements at each of the $N$ delay steps; the in-phase component
(like in the real correlator) and also the quadrature component. The
quadrature component is detected by inserting a $-90\dgr$ phase shift
into one arm. For the quadrature component, the cosine in the
cross-correlation function (Eq.~(\ref{eqn:corr-response1})) is
replaced with a sine;
\begin{equation}
R_2 = R_0 \frac{\sin \pi \Dnu \Dtau}{\pi \Dnu \Dtau}
\sin \bigl[ 2\pi \bigl(\nuLO \taug \pm \nuIF \Dtau 
+ \phiLO \bigr) \bigr]. \label{eqn:corr-response2} 
\end{equation}
\revd{We have also substituted in the residual delay $\Dtau$ defined
  in Eq.~(\ref{eqn:Dtau}) for clarity.} The complex correlator samples
the in-phase and quadrature components at $N$ delay steps. The two
components can be conveniently expressed by a sum of the two
orthogonal signals,
\begin{equation}
\Rx = R_1 + j R_2\label{eqn:corr-response3}
\end{equation}
or written as an exponential;
\begin{equation}
\Rx = R_0 \frac{\sin \pi \Dnu \Dtau}{\pi \Dnu \Dtau} 
\exp \bigl[ 2\pi \imagj \bigl(\nuLO \taug \pm \nuIF \Dtau 
+ \phiLO \bigr) \bigr]. \label{eqn:corr-response4}
\end{equation}
The delay steps of the Fourier transform correlator are determined by
the total bandwidth of the correlator. For the critically-sampled real
correlator, the delays are in steps of $T = 1 / (2 \Dnu$). The complex
correlator samples at only half-Nyquist rate ($T = 1 / \Dnu$) because
the cross-power spectrum of a complex cross-correlation function is
single-sided (see Fig.~\ref{fig:ccf-spec}b).

\subsection{Fringe-rotation}

Conventional fringe-rotation transforms the data so that the phase
centre maps to the tracking centre. This is done by compensating for
the cosine component of Eq.~(\ref{eqn:corr-response1}). Fringe-rotation
can be applied either in hardware by controlling the phase of the
local oscillator ($\phiLO$) or in software after recording the
data. We will assume the latter case. The data must be sampled
fast enough to avoid fringe-washing. For the complex correlator, the
fringe data (Eq.~(\ref{eqn:corr-response4})) is multiplied by
\begin{equation}
\Ffull = \exp \bigl[ -2\pi \imagj \bigl(\nuLO \taug \pm \nuIF \Dtau 
\bigr) \bigr]. \label{eqn:frot-full}
\end{equation}
This fringe-rotation factor can be calculated exactly from a
combination of the geometry of the baseline (antenna pair), the
position of the source and the frequencies $\nuLO$ and
$\nuIF$. Fringe-rotation will stop the phase-wrapping of the the
fringes to give
\begin{equation}
\Rx \Ffull = R_0 \frac{\sin \pi \Dnu \Dtau}{\pi \Dnu \Dtau} 
\exp \bigl[ 2\pi \imagj \phiLO \bigr]. \label{eqn:froted1}
\end{equation}
The Fourier transform of the stopped fringes gives the spectral data
or visibilities. The visibilities can then be used to map the source.

For the real correlator (Eq.~(\ref{eqn:corr-response1}) and as
illustrated by Fig~\ref{fig:ccf-spec}a), we cannot directly rotate the
fringes in software because we do not have both the amplitude and
phase information at each delay step. So the data must first be Fourier
transformed across the delays into spectral channels (which is the
usual operation for Fourier transform correlators). Each spectral
channel is then fringe-rotated. The data from the
complex correlator can also be fringe-rotated after the Fourier
transform. This may be advantageous, for example when the fractional
bandwidth ($\Dnu / \nuRF$) is high and the centre frequency used for
fringe rotation is less constrained.

\subsection{Spectral synthesis and mapping}\label{sec:spec-synth-map}

In a conventional signal processing pipeline, the data recorded at
each time sample are Fourier transformed across the delays to give the
spectral data at each sub-band. For example, the continuous Fourier
transform of Eq.~(\ref{eqn:froted1}) gives a rectangular spectrum as
expected. The amplitude is calibrated against a flux calibrator source
and corrected for changes in the system temperature (including effects
like airmass). The phase is also calibrated against a phase calibrator
source. This will remove phase offsets like $\phiLO$. In the ideal
case, an unresolved source at the tracking centre will produce
visibility data with a constant amplitude and zero phase. The
successive visibility samples are time-averaged to reduce the data
load and gridded to the two-dimensional aperture plane. This gives the
spatial frequency components of the map. The coordinates of the
visibility data in the aperture plane depends on the baseline geometry
relative to the source and the RF of the sub-band. When all the
visibilities have been gridded and appropriately weighted, the
aperture plane is Fourier transformed to the map plane. The result is
the dirty map -- the convolution of the map with the synthesised
beam. The synthesised beam of an interferometer usually has large
sidelobes so the dirty map can be deconvolved using standard
techniques such as CLEAN. Parts of the steps for reducing the data are
summarised in Fig.~\ref{fig:flowchart}.

\section{Oversampling method}\label{sec:os-method}

As we already showed in Sect.~\ref{sec:intro}, it is possible to
reduce aliasing by oversampling the cross-correlation function above
the the Nyquist rate. This inserts a buffer zone in the spectral
domain between the signal spectrum and the images. Using the new
method that we will describe here, we could achieve this without
increasing the number of detectors in the correlator. 

\revd{The source is tracked as usual and the path differences between
  the antennas are compenstated in discrete steps using path
  compensators. Normally, the data are fringe-rotated to align the
  phase centre with the tracking centre. This {\it stops} the
  fringes. Instead, we apply a modified fringe-rotation
  (Sect.~\ref{sec:mod-frot}) and allow the fringes to vary. When the
  path compensator steps, the data since the last step are collected
  together. The varying fringes trace out the cross-correlation
  function in Eq.~\ref{eqn:corr-response4} with $\taug = 0$. The
  function is densely sampled so the Fourier transform
  (Sect.~\ref{sec:spec-synth}) introduces negligible
  aliasing\footnote{The oversampling method shares some analogies with
    drift scans. In a drift scan, the source is tracked but the path
    compensator is kept fixed. The spectrum can then be estimated by
    taking the subsequent Fourier transform of the time-series
    data.}.}

\subsection{Modified fringe-rotation}\label{sec:mod-frot}

The aim of the method is to build up an oversampled set of data points
for the cross-correlation function in {$\Dtau$}. But this is
complicated because the exponential term in the response of the
complex correlator (Eq.~(\ref{eqn:corr-response4})) is also dependent
on $2\pi\nuLO \taug$. It is possible to remove just this term from the
equation by applying a modified fringe-rotation function instead;
\begin{equation}
\Ffr = \exp \bigl[ -2\pi \imagj \nuLO \taug \bigr].
\label{eqn:frot-sr}
\end{equation}
The resulting fringe-rotated data will be a function of $\Dtau$;
\begin{equation}
\Rx \Ffr = R_0 \frac{\sin \pi \Dnu \Dtau}{\pi \Dnu \Dtau} 
\exp \bigl[ 2\pi \imagj \bigl( \pm \nuIF \Dtau + \phiLO \bigr) \bigr]. 
\label{eqn:froted2}
\end{equation}
By collecting consecutive time samples, we can build up an oversampled
measurement of the cross-correlation function. We assume that the
underlying signal is not changing over the time it takes to sample the
whole cross-correlation function and we will return to this issue in
Sect.~\ref{sec:spec-synth}.

In contrast, the real correlator is not very well suited to the
oversampling method. The fringes cannot be directly fringe-rotated so
it first needs to be Fourier transformed\footnote{In signal
processing, window functions are often applied to reduce
sidelobes. But a window function should not be applied at this stage
because the spectral resolution needs to be as narrow as possible to
avoid aliasing -- aliasing from the overlapping main lobes is a more
acute problem.}. The spectral channels are fringe-rotated by the
modified factor\footnote{The conjugate relationship between the
positive and negative spectra in a real correlator means that when
applying fringe-rotation to a real correlator, the conjugate of
Eq.~(\ref{eqn:frot-sr}) may have to be applied to one-half of the
spectrum} $\Ffr$ from Eq.~(\ref{eqn:frot-sr}) as before and then
inverse Fourier transformed back\revd{;} 
\begin{equation}
{R_1}^\prime = \mathrm{FT}^{-1}\bigl[
  \Ffr \, \mathrm{FT}\bigl[R_1(\Dnu, \Dtau, \taug)\bigr]\bigr].
\label{eqn:froted-real}
\end{equation} 
\revd{Note that this gives the rotated fringes as a function of
  $\Dtau$.}  In addition to the computational burden, the Fourier
transform step will cause aliasing. We will investigate this
phenomenon through simulations in Sect.~\ref{sec:results-ideal-corr}.
This shortcoming for the real correlators may be overcome with a
hardware fringe-rotation system. This approach also has the benefit of
slowing down the fringes so that the data can be sampled at a lower
rate. If fringe-rotation is to be applied in software, complex
correlators are preferable because direct fringe-rotation is faster
and much cleaner. \revd{However broadband {--90\dgr} phase shifters
  can be difficult to design.}

So far we have assumed an ideal square passband. In reality, the
source's spectrum may not be flat and the instrument may have a
sloping passband. These factors may displace the centre-frequency and
degrade the system's sensitivity. But following methods outlined by
\citet{thompson1982,thompson2001}, we calculated that the losses from
fringe-rotation are negligible compared to losses from the sloping
spectrum.


\subsection{Spectral synthesis}\label{sec:spec-synth}

\begin{figure}[hbtp]
  \centering 
  \includegraphics[width=3.5in]{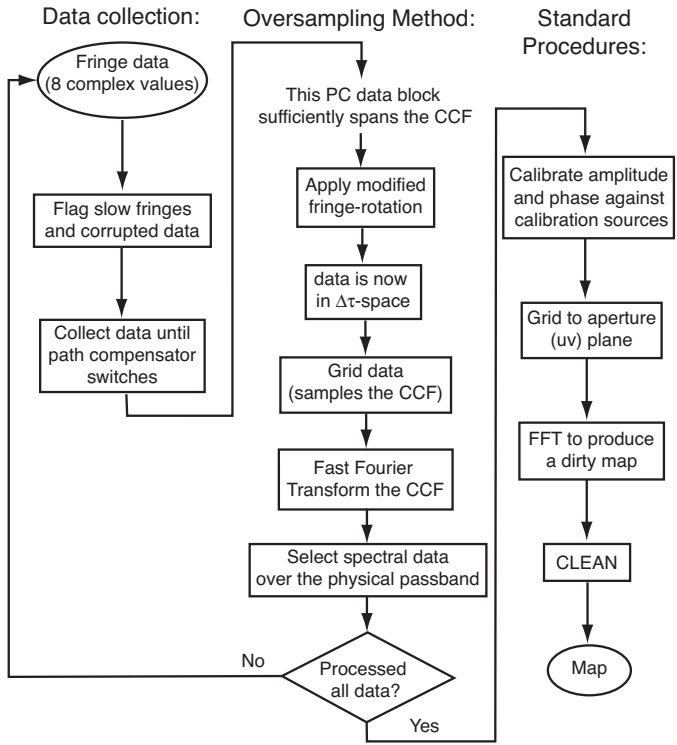}
\caption{A flowchart summarising the oversampling method.}
\label{fig:flowchart}
\end{figure}

We now have samples of the fringe-rotated cross-correlation function
against {$\Dtau$}. We now divide the time-series data between path
compensation changes so that in each \textit{PC data block}, the
cross-correlation function is completely sampled over the required
range in {$\Dtau$}. The PC data block is then gridded and each pixel
is weighted by the system temperature at the corresponding sample
time. The gridded PC data block is then Fourier transformed\footnote{A
window function could be applied to the gridded data before the
Fourier transform step but it needs to be chosen with care; we
discussed in {\paperone} that window functions could systematically
bias the estimated flux density of sources that are not at the
tracking centre.} to give the oversampled spectrum. The wide spectral
buffer between the signal and the images reduces aliasing. The
synthesised spectrum over the passband is extracted, calibrated and
gridded to the aperture ($uv$) plane as described before. This process
is repeated for each PC data block. We have summarised the steps as a
flowchart in Fig.~\ref{fig:flowchart} and we will describe some examples
through simulations in the next section.

The period of the PC data block depends on the baseline and will vary
over the course of an observation. A fundamental requirement is that
the PC data block period needs to be short enough to avoid
time-average smearing \citep[see][]{thompson2001}. So the periods of
the PC data blocks {\Tpcb} must satisfy
\begin{equation}
\Tpcb < (d / D) / \OmegaE,
\end{equation}
where $d$ is the antenna diameter, $D$ is the baseline and {\OmegaE}
is the angular speed of the Earth's rotation. The period of a PC data
block for a complex correlator is given by:
\begin{equation}
\Tpcb = (\nuRF / \Dnu) / \nuf,
\end{equation}
where {\nuf} is the fringe frequency. For real correlators, where the
shortest path compensation bit is shorter, {\Tpcb} is half the above
value. In AMI, when the fringe rate drops below $\nuf \sim
10^{-3}\,$Hz, it becomes difficult to separate the signal from
slowly-varying non-astronomical noise, so these samples are usually
rejected. Some additional data will have to be rejected at low fringe
rates because they cannot completely fill the PC data block. But the
proportion of lost data is small for a telescope like AMI.

%
%
%
%
%
%
%
%
%
%
%
%
%
%
%

\section{Simulations}\label{sec:simulations}

We tested the oversampling concept with simulations of a model
telescope based on AMI. The observing frequency is 15{\GHz} with a
6{\GHz} bandwidth, mixed down with an LO of 24{\GHz} to give an IF
band of 6--12{\GHz}. Path compensation is applied in the IF at
discrete steps. The shortest delay line is the same as the
correlator's delay steps so that the gaps in the cross-correlation
function can be filled over the course of a single path
step. Fringe-rotation is applied in software, rather than in hardware
so the fringes need to be sampled sufficiently fast. For a relatively
short $5\,$m east-west baseline, we expect fringe periods longer than
$50\,$s, so sampling at $1\,$Hz is more than sufficient. 8 spectral
channels are synthesised from the full 6{\GHz} bandwidth. We
investigated both the real and complex correlators. The real
correlator has 16 delay steps and the complex correlator makes a pair
of measurements at each of its 8 delay steps. The parameters of the
model telescope are summarised in Table~\ref{tab:tel-param}.

\begin{table}[hbtp]
\caption{Model telescope parameters}             
\label{tab:tel-param}      
\centering          
\begin{tabular}{l c}     
\hline\hline       
   RF       & 15{\GHz} (12--18{\GHz})\\
Bandwidth   & 6{\GHz}\\
Synthesised sub-bands & 0.75{\GHz}\\
   LO (Lower sideband reception) & 24{\GHz}\\
   IF       & 6--12{\GHz}\\
Dish diameter & $3.7\,$m\\
Baseline range & 5--$20\,$m\\
Simulated baseline    & 5$\,$m east-west (250$\lambda$)\\
Integration       & 1$\,$s\\
Spectral channels & 8 (0.75{\GHz} each)\\
Latitude    & 52\dgr\\
Test source declination & 52\dgr\\
\hline
\textit{Real correlator:}\\
Correlator delay steps & 16 samples\\
Correlator delay steps & $83\,$ps (25$\,$mm)\\
Path compensator steps & $83\,$ps (25$\,$mm)\\
Tracking hour angle range   & $\pm 56\,$s\\
\hline                  
\textit{Complex correlator:}\\
Correlator delay steps & 8 complex samples\\
Correlator delay steps & $167\,$ps (50$\,$mm)\\
Path compensator steps & $167\,$ps (50$\,$mm)\\
Tracking hour angle range    & $\pm 112\,$s\\
\hline\hline     
\end{tabular}
\end{table}

\subsection{Simulations method}

\begin{figure*}[htbp]
  \centering 
  \mbox{\subfigure[Real correlator]{
  \includegraphics[width=3.45in]{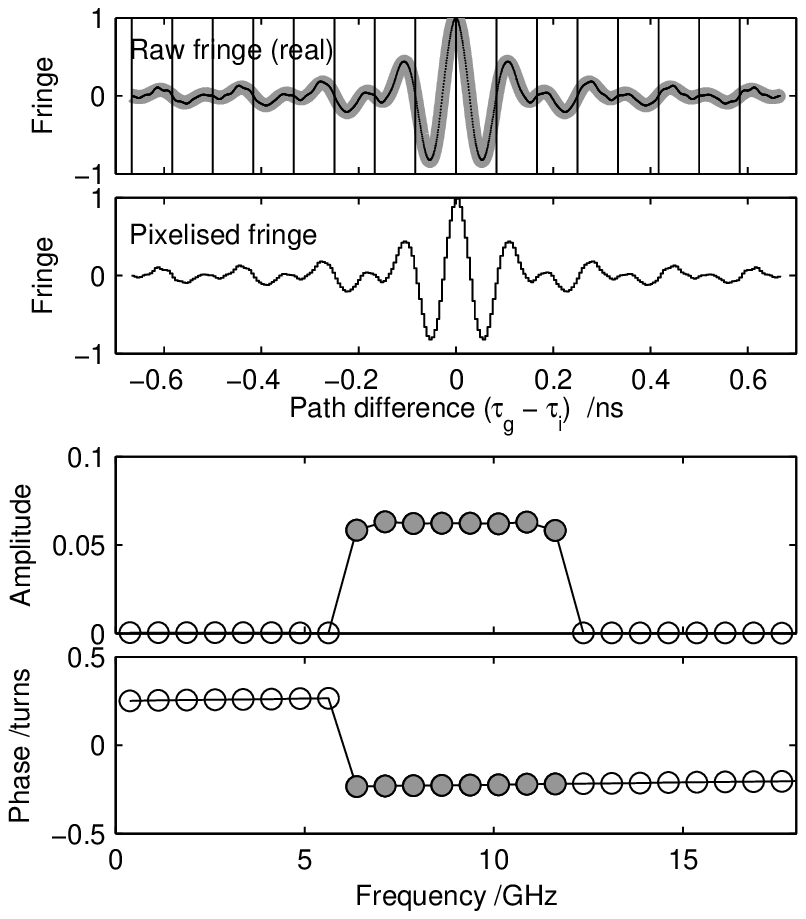} }}
  \mbox{\subfigure[Complex correlator]{
  \includegraphics[width=3.45in]{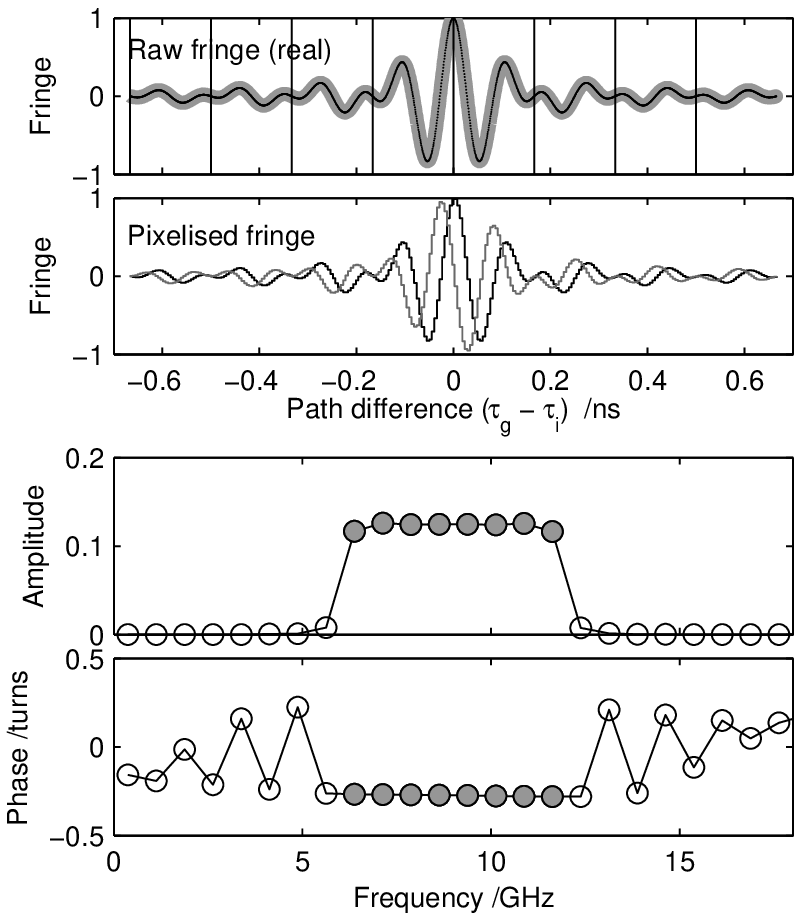} }} 
  \caption{The oversampled spectral synthesis for a real correlator
    {\bf a)} and complex correlator {\bf b)}. The top plots are the
    real components of the fringes. (1) The thick grey lines mark the
    ideal cross-correlation function we would expect \revd{from
      Eq.~(\ref{eqn:corr-response1}) for {\bf a)} and the real
      component of Eq.~(\ref{eqn:corr-response4}) for {\bf b)}. The
      geometrical delay {\taug} for the thick grey curves are set to
      0.} The dark dots (seen as continuous curves at this resolution)
    are the \revd{modified-fringe-rotated data points} measured by the
    correlator at each time sample \revd{(from
      Eq.~(\ref{eqn:froted-real}) and Eq.~(\ref{eqn:froted2})
      respectively)}. \revd{In {\bf b)}, the imaginary components have
      not been shown for clarity.}  The vertical lines indicate the
    range in {\Dtau} spanned by each delay; the curve in each section
    is traced out from the left boundary to the right. (2) In the
    second plot, the \revd{modified-fringe-rotated} data points are
    pixelised into 256 pixels. For the complex correlator, the dark
    curve is the real component \revd{of Eq.~(\ref{eqn:froted2})} and
    the grey curve is the imaginary component. (3) The pixelised data
    are fast Fourier transformed. The bottom two plots show the
    amplitude and phase of the spectrum. The estimated spectrum
    extends to higher frequencies so only a portion is shown. The
    filled circles mark the synthesised spectral channels of interest
    over the physical passband of the system.\label{fig:superres}}
\end{figure*}

We simulate the fringes measured at each delay step using
Eq.~(\ref{eqn:corr-response1}) and (\ref{eqn:corr-response4}). We
modelled an unresolved source at the centre of the field but the
method will work for any sources because we are simply measuring the
cross-correlation function. The antennas track the point source about
the transit (when the source is due south) over an hour angle range of
$\pm 56\,$s for the real correlator or $\pm 112\,$s for the complex
correlator. This fills the CCF function over a range of $\Dtau = \pm
0.61\,$ns. The time-stream data is then fringe-rotated as outlined in
Sect.~\ref{sec:mod-frot}. The cross-correlation function is now a
function of $\Dtau$ and we grid the data points to 256 pixels (see
the upper set of plots in Fig.~\ref{fig:superres}). We now have an
oversampled measurement of the cross-correlation function. If the
pixels are too narrow, some pixels may not contain any data points and
this will degrade the spectrum. At the same time, the pixels must be
fine enough so that the cross-correlation function is not washed
out. Individual pixels could be weighted by factors such as the number
of data points that contribute to the pixel and the system temperature
associated with the data points. Finally, we apply the fast Fourier
transform (FFT) to the gridded data and extract the signal spectrum
over the {6--12\GHz} passband (lower plots in
Fig.~\ref{fig:superres}). The amplitude and phase can be calibrated
against an astronomical source by standard methods.

\subsection{Results and discussions}

\subsubsection{Ideal correlator}\label{sec:results-ideal-corr}

The recovered spectrum is in steps of {0.75\GHz} and is the same as
when the conventional {\it one-shot} method is used. The oversampled
spectrum extends from $\nuIF =$~DC to {104\GHz} and only a portion is
shown in Fig.~\ref{fig:superres}. The spectrum for the complex
correlator in Fig.~\ref{fig:superres}a does not appear to suffer from
aliasing.  The arguments in Sect.~\ref{sec:intro}, together with
Fig.~\ref{fig:under-vs-oversamp} suggests that the oversampling method
should reduce aliasing.

\begin{figure}[bp]
  \centering
  \includegraphics[width=3.45in]{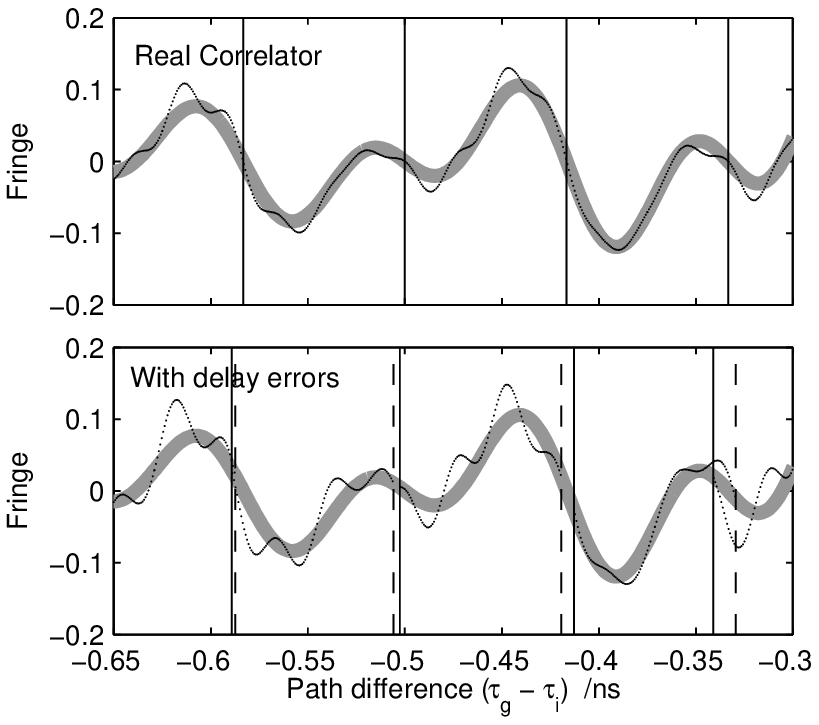}
  \caption{Details of the reconstructed cross-correlation function for
    the real correlator \revd{in Fig.~\ref{fig:superres}a towards the
      edge of the delay range. The thick grey curve is the expected
      function from Eq.~(\ref{eqn:corr-response1}) with $\taug$ set to
      0. The dots are the data points generated from
      Eq.~(\ref{eqn:froted-real}) after the modified fringe rotation.}
    The upper plot is for a perfect real correlator (from
    Fig.~\ref{fig:superres}a) and the lower plot is with delay errors
    (from Fig.~\ref{fig:lag-err-noise}a). The solid vertical lines
    indicate the position of each delay in {\Dtau} at the beginning of
    the PC data block. The dashed vertical lines are the position at
    the end of the PC data block. Delay errors lead to missing samples
    or overlapping samples inside the narrow gaps indicated by these
    lines.  Fringe-rotation aliasing is apparent in both cases and it
    is also worse when there are delay errors.\label{fig:zoom-in}}
\end{figure}

In the real correlator, fringe-rotation introduces aliasing via the
Fourier transform. This can be seen as small cycles in the data points
(Fig.~\ref{fig:superres} and details in Fig.~\ref{fig:zoom-in}). These
arise from the time-domain alias cycles in
Fig.~\ref{fig:alias-cycles}. Fringe-rotation aliasing is worse near
the edge of the cross-correlation function. The root mean square (RMS)
deviation of the recovered spectrum from the ideal case is negligible
(below 0.1~percent). But if there are delay errors, the RMS deviation
may become significant and this will be discussed in
Sect.~\ref{sec:delay-err}. If using the oversampling method on real
correlators, we would recommend hardware fringe-rotation. If software
fringe-rotation is chosen, as shown here, the effects of
fringe-rotation aliasing needs to be investigated further. Gridding is
equivalent to down-sampling and the conventional practice of
introducing a lowpass anti-aliasing filter before gridding may be
beneficial. With larger fractional bandwidths ($\Dnu / \nuRF$), the
small cycles from fringe-rotation aliasing will become longer relative
to the pixel size.

\subsubsection{Correlators with delay errors}\label{sec:delay-err}

The oversampling method also corrects for delay errors prevalent in
analogue Fourier transform correlators. It is possible to determine
the delay errors either by bench measurements or by tracking a source
with the path compensator held fixed (see \citealt{harris2001};
{\paperone}). The delay errors are incorporated into calculating
$\Dtau$ and gridded uniformly so the recovered frequency channels are
also uniformly spaced.

Figures~\ref{fig:lag-err-noise}a and \ref{fig:lag-err-noise}b show
simulations for both the real and complex correlators with delay
errors of up to 10~percent of the delay steps. For the complex
correlator, we assumed that the delay errors for each pair of
detectors are the same. This is a reasonable assumption for our
correlator architecture because we found that most delay errors arise
in the long delay lines common to both the real and imaginary signals
(see {\paperone}). In both correlators, there are gaps in the samples
from the irregular delay steps. The recovered spectrum is given by the
spectral response of this distorted window function convolved with the
true spectrum. This will inevitably worsen the inter-channel spectral
leakage.

The presence of delay errors makes aliasing \revd{resulting from
  fringe-rotation of the real correlator data} noticeably worse
(Fig.~\ref{fig:lag-err-noise}a and details in
Fig.~\ref{fig:zoom-in}). This is consistent with our previous findings
(\paperone). When fringe-rotating the spectral channels from the real
correlator, we assumed regularly-spaced centre-frequencies at the
design frequencies. These may not be the best choice for an
irregularly-spaced correlator and may even introduce small phase
errors during fringe-rotation. Fringe-rotation aliasing gives rise to
a low-level spread-spectrum noise in the amplitude plot
(Fig.~\ref{fig:lag-err-noise}a). Perhaps the most significant evidence
against applying software fringe-rotation to a real correlator is the
potential for systematic errors.We found about 3~percent RMS deviation
from the ideal spectrum when there are delay errors. The alias cycles
seem to be largely responsible for this because the RMS deviation for
a complex correlator with lag errors was less than 0.01~percent.

It is worth noting that a correlator with delay errors will
fundamentally reduce sensitivity and no amount of signal processing
can compensate for it. This is because the SNR of the recovered
spectrum is optimised when sampled at Nyquist steps (or at
half-Nyquist steps for complex correlators). Delay errors cause the
noise components between the delay steps to be correlated, so the
measurements are no longer independent and the SNR is degraded. The
benefits of applying the oversampling method in the presence of delay
errors are: (1) The recovered spectral channels are regularly gridded
and are at the designed centre frequencies. (2) Normally, delay errors
worsen the effects of aliasing (\paperone).
Fig.~\ref{fig:lag-err-noise}b and the RMS figure for the complex
correlator suggest that the oversampling method reduces aliasing, even
in the presence of delay errors.

\subsubsection{In the presence of noise}\label{sec:add-noise}

\begin{figure*}[p]
  \centering 
  \mbox{\subfigure[Real correlator with delay errors]{
  \includegraphics[width=3.44in]{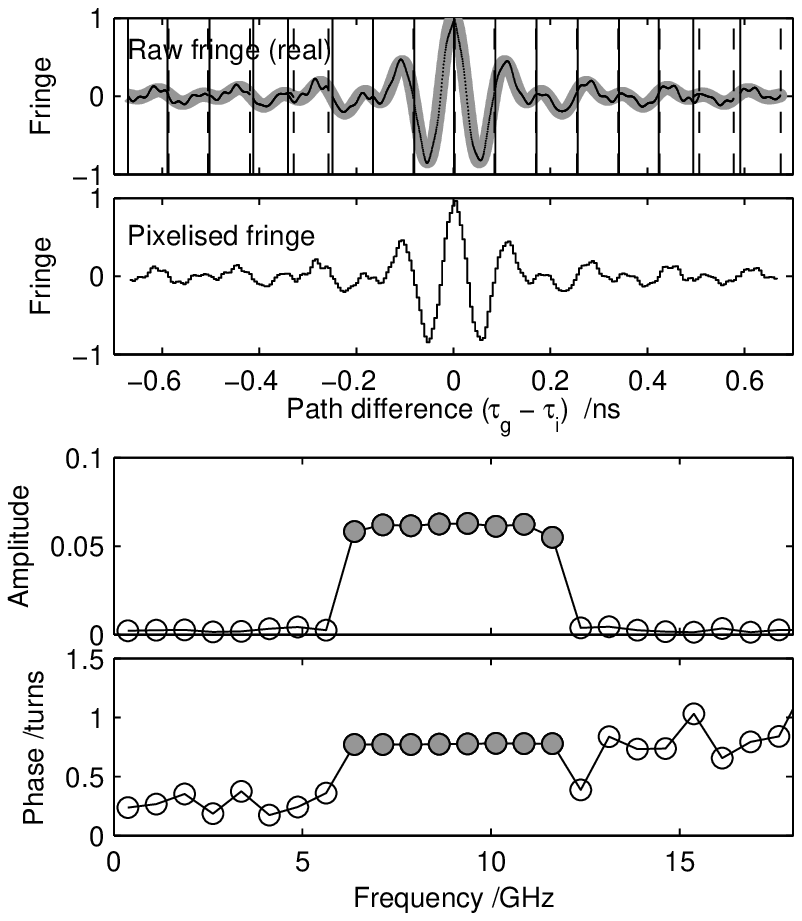} }}
  \mbox{\subfigure[Complex correlator with delay errors]{
  \includegraphics[width=3.44in]{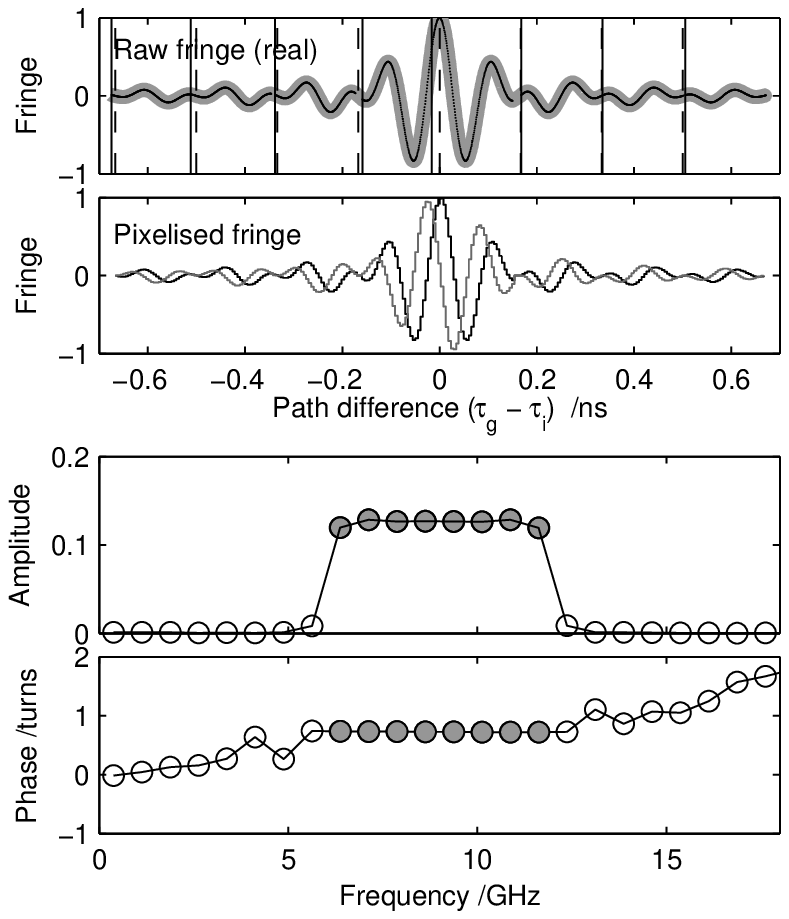}}} 
  \mbox{\subfigure[Complex correlator with SNR = 1]{
  \includegraphics[width=3.44in]{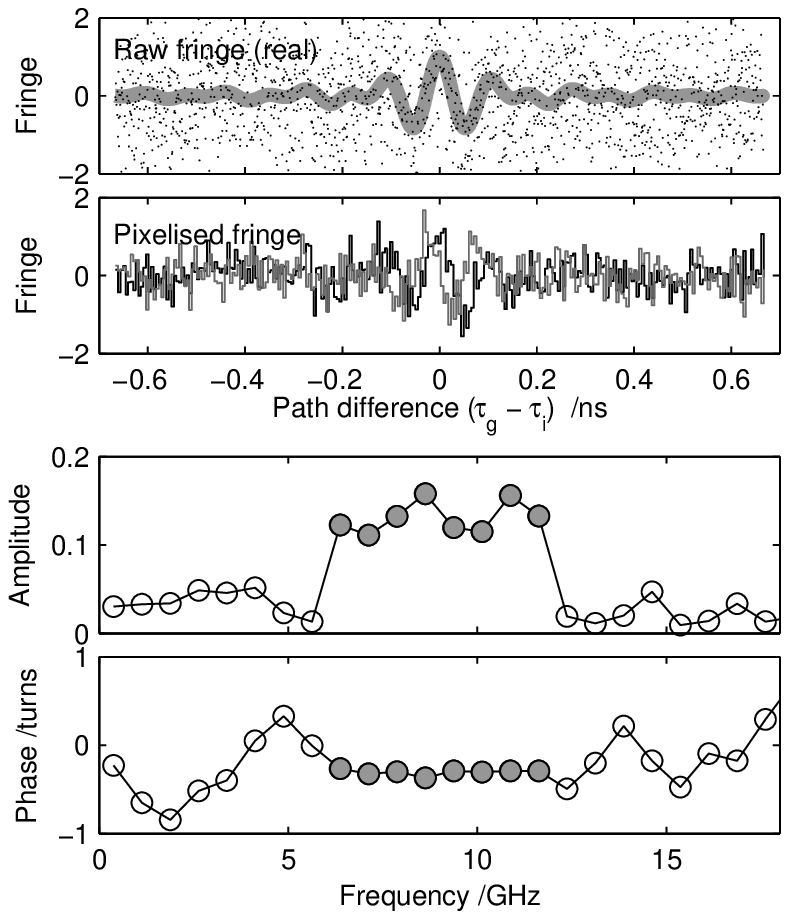} }}
  \mbox{\subfigure[Complex correlator with SNR = 0.25 (stacking 16 data sets)]{
  \includegraphics[width=3.44in]{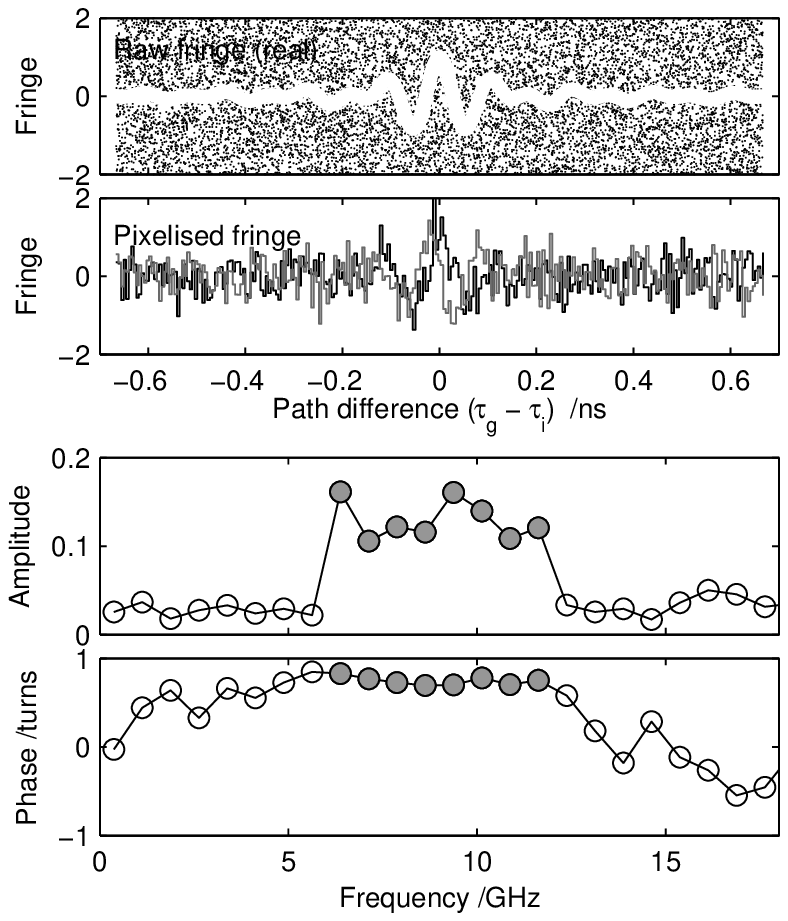} }} 
  \caption{{\bf a)} and {\bf b)}: The oversampling method in the
    presence of delay errors. In both cases, the errors are up to
    10~percent of the delay steps and there are \revd{visible} gaps in
    the data. The solid vertical lines mark the left-most range of
    each delay and the dashed vertical lines mark the right-most
    range. The delay errors can be clearly seen from the relative
    alignment of these lines. The other effect in the real correlator
    {\bf a)} is the higher level of fringe-rotation aliasing in the
    cross-correlation function (see details in
    Fig.~\ref{fig:zoom-in}). \revd{The thick grey curves are the ideal
      cross-correlation curves from Eq.~(\ref{eqn:corr-response1}) with
      {\taug} set to 0. The dark points are the
      modified-fringe-rotated data points generated from
      Eq~(\ref{eqn:froted-real}) and Eq.~(\ref{eqn:froted2}).} {\bf
      c)} and {\bf d)}: Gaussian random noise was added to the fringes
    of a complex correlator. The instantaneous SNR is quantified by
    the ratio of the peak of the cross-correlation function to the
    standard deviation of the noise. In {\bf d)}, 16 data sets are
    stacked so that the recovered spectrum has the same SNR as {\bf
      c)}. The phases have been unwrapped
    arbitrarily\label{fig:lag-err-noise}}
\end{figure*}

In Figs.~\ref{fig:lag-err-noise}c and \ref{fig:lag-err-noise}d, we
added independent gaussian random noise to the fringes of a complex
correlator. Noise in each time-frame are independent and it can be
shown that in our idealised system, the noise between delay steps are
also independent. The recovered spectra are noisier as expected but the
oversampling method is robust in the presence of noise. It is also a
linear process so successive data sets can be stacked
(Fig.~\ref{fig:lag-err-noise}d). Because the noise is independent
between samples and also between the delay steps, the noise energy is
spread over the whole of the oversampled spectrum. We only select
spectral channels over the passband and discard the rest. This does
not improve the SNR beyond what can be achieved by standard
methods. The SNR improvement from applying this matched filter is the
same as from averaging successive samples in the standard methods.

\subsection{Issues in a practical system}\label{sec:practical-issues}

Several issues will need to be addressed when using the oversampling
method in a practical system and design tolerances must be
set. Firstly, more detailed simulations of the method should be
conducted to quantify the reduction in aliasing. So far, we have
assumed that the passbands of the detectors at each delay steps are
identical. Any variation will restrict the effective bandwidth and
degrade the sensitivity. If there are variations in the passbands,
each detector will trace out a different cross-correlation function
because the cross-correlation function is related to the Fourier
transform of the passband. When the data points are merged to
reconstruct the full cross-correlation function, there will be
discontinuities between the detector outputs. However the degradation
is probably comparable to standard methods.

If the path compensators do not step exactly in multiples of one delay
step (due to manufacturing tolerances in the path compensators), there
will be unsampled gaps in {\Dtau}. Optimising the path compensation
for an array can be more complicated and some delay ranges ($\Dtau$)
may not be fully sampled. Alternatively, some samples may have to be
rejected because they are corrupted. These missing samples will
degrade the overall sensitivity in the same way as delay
errors. Making the grid coarser may mitigate some of this but at the
expense of washing out the cross-correlation. The data needs to be
stable over the PC data block period. This could be worked round by
opting for real correlators \revd{(with hardware fringe-rotation)},
which cuts down {\Tpcb} by a half.


\subsection{Application to AMI}

Although both the real and complex correlators were trialled for AMI,
the real correlator had been chosen (see {\paperone}). As demonstrated
here, the oversampling method is not suited for real correlators
relying on software fringe-rotation. Later on, it was found that the
bottom two frequency channels in AMI (12--13.5{\GHz}) suffered from
geostationary satellite interference. The effects became noticeable
when observing at low declinations and it was decided to block this
frequency range all together with hardware filters. Although this
reduces the bandwidth of the instrument, it also goes some way towards
reducing the aliasing on one side of the spectrum. This change avoided
the need for the more complicated oversampling method and
subsequently, we did not implement it in the AMI data reduction
pipeline.

\section{Conclusions}

We have described a software-based method for overcoming aliasing in
critically-sampled Fourier transform correlators. The method reduces
aliasing by reconstructing the oversampled cross-correlation function
from successive samples. Data from complex correlators can be
fringe-rotated directly and efficiently in software. If using this
method on real correlators, the signal should be fringe-rotated in
hardware to avoid systematic errors.

The edge spectral channels are usually discarded because they are the
worst affected by aliasing. For an 8-channel complex correlator
simulated here, the sensitivity could be improved by up to 15~percent
by retaining the edge channels. The additional computational overhead
and complexity are offset by an improvement in sensitivity and
reduction in systematic errors from aliasing. This could be a
significant benefit for Fourier transform correlators with less than
10 spectral channels. Some data with low fringe rates will have to be
rejected, but for AMI-like telescopes with fractional bandwidths $\Dnu
/ \nuRF \gtrsim 0.4$ and maximum baseline to dish ratios of $D/d
\lesssim 10$, the fraction of data lost is relatively small. We have
also shown that we can naturally compensate for delay errors to
recover regularly-gridded frequency channels at the design
frequencies.


\begin{acknowledgements}
The authors would like to thank Peter Duffet-Smith and
Christian Holler for helpful comments on the manuscript. We would like
to add our special thanks to the referee for many helpful suggestions.
\end{acknowledgements}

\bibliographystyle{aa}  
\bibliography{8117kaneko}   

\end{document}